\PassOptionsToPackage{unicode}{hyperref}
\PassOptionsToPackage{hyphens}{url}
\PassOptionsToPackage{dvipsnames,svgnames,x11names}{xcolor}
\documentclass[
  letterpaper,
  DIV=11,
  numbers=noendperiod]{scrartcl}

\usepackage{amsmath,amssymb}
\usepackage{iftex}
\ifPDFTeX
  \usepackage[T1]{fontenc}
  \usepackage[utf8]{inputenc}
  \usepackage{textcomp} 
\else 
  \usepackage{unicode-math}
  \defaultfontfeatures{Scale=MatchLowercase}
  \defaultfontfeatures[\rmfamily]{Ligatures=TeX,Scale=1}
\fi
\usepackage{lmodern}
\ifPDFTeX\else  
\fi
\IfFileExists{upquote.sty}{\usepackage{upquote}}{}
\IfFileExists{microtype.sty}{
  \usepackage[]{microtype}
  \UseMicrotypeSet[protrusion]{basicmath} 
}{}
\makeatletter
\@ifundefined{KOMAClassName}{
  \IfFileExists{parskip.sty}{%
    \usepackage{parskip}
  }{
    \setlength{\parindent}{0pt}
    \setlength{\parskip}{6pt plus 2pt minus 1pt}}
}{
  \KOMAoptions{parskip=half}}
\makeatother
\usepackage{xcolor}
\ifx\paragraph\undefined\else
  \let\oldparagraph\paragraph
  \renewcommand{\paragraph}[1]{\oldparagraph{#1}\mbox{}}
\fi
\ifx\subparagraph\undefined\else
  \let\oldsubparagraph\subparagraph
  \renewcommand{\subparagraph}[1]{\oldsubparagraph{#1}\mbox{}}
\fi

\usepackage{longtable,booktabs,array}
\usepackage{calc} 
\usepackage{etoolbox}
\makeatletter
\patchcmd\longtable{\par}{\if@noskipsec\mbox{}\fi\par}{}{}
\makeatother
\IfFileExists{footnotehyper.sty}{\usepackage{footnotehyper}}{\usepackage{footnote}}
\makesavenoteenv{longtable}
\usepackage{graphicx}
\makeatletter
\def\maxwidth{\ifdim\Gin@nat@width>\linewidth\linewidth\else\Gin@nat@width\fi}
\def\maxheight{\ifdim\Gin@nat@height>\textheight\textheight\else\Gin@nat@height\fi}
\makeatother
\setkeys{Gin}{width=\maxwidth,height=\maxheight,keepaspectratio}
\makeatletter
\def\fps@figure{htbp}
\makeatother
\newlength{\cslhangindent}
\newlength{\csllabelwidth}
\newlength{\cslentryspacingunit} 
\newenvironment{CSLReferences}[2] 
 {
  \setlength{\parindent}{0pt}
  \ifodd #1
  \let\oldpar\par
  \def\par{\hangindent=\cslhangindent\oldpar}
  \fi
  \setlength{\parskip}{#2\cslentryspacingunit}
 }%
 {}
\usepackage{calc}

\usepackage{booktabs}
\usepackage{caption}
\usepackage{longtable}
\usepackage{colortbl}
\usepackage{array}
\usepackage{anyfontsize}
\usepackage{multirow}
\KOMAoption{captions}{tableheading}
\makeatletter
\@ifpackageloaded{caption}{}{\usepackage{caption}}
\AtBeginDocument{%
\ifdefined\contentsname
  \renewcommand*\contentsname{Table of contents}
\else
  \newcommand\contentsname{Table of contents}
\fi
\ifdefined\listfigurename
  \renewcommand*\listfigurename{List of Figures}
\else
  \newcommand\listfigurename{List of Figures}
\fi
\ifdefined\listtablename
  \renewcommand*\listtablename{List of Tables}
\else
  \newcommand\listtablename{List of Tables}
\fi
\ifdefined\figurename
  \renewcommand*\figurename{Figure}
\else
  \newcommand\figurename{Figure}
\fi
\ifdefined\tablename
  \renewcommand*\tablename{Table}
\else
  \newcommand\tablename{Table}
\fi
}
\@ifpackageloaded{float}{}{\usepackage{float}}
\floatstyle{ruled}
\@ifundefined{c@chapter}{\newfloat{codelisting}{h}{lop}}{\newfloat{codelisting}{h}{lop}[chapter]}
\floatname{codelisting}{Listing}

\makeatother
\makeatletter
\@ifpackageloaded{caption}{}{\usepackage{caption}}
\@ifpackageloaded{subcaption}{}{\usepackage{subcaption}}
\makeatother
\makeatletter
\@ifpackageloaded{tcolorbox}{}{\usepackage[skins,breakable]{tcolorbox}}
\makeatother
\makeatletter
\@ifundefined{shadecolor}{\definecolor{shadecolor}{rgb}{.97, .97, .97}}
\makeatother
\ifLuaTeX
  \usepackage{selnolig}  
\fi
\IfFileExists{bookmark.sty}{\usepackage{bookmark}}{\usepackage{hyperref}}
\IfFileExists{xurl.sty}{\usepackage{xurl}}{} 
\hypersetup{
  pdftitle={Do LLMs Act as Repositories of Causal Knowledge?},
  pdfauthor={Nick Huntington-Klein; Eleanor J. Murray},
  colorlinks=true,
  linkcolor={blue},
  filecolor={Maroon},
  citecolor={Blue},
  urlcolor={Blue},
  pdfcreator={LaTeX via pandoc}}

\title{Do LLMs Act as Repositories of Causal Knowledge?}
\author{Nick
Huntington-Klein\thanks{Thanks to Phuong Nguyen and Meet Panjwani for research assistance.} \and Eleanor
J. Murray}
\date{}

\begin{document}
\maketitle
\begin{abstract}
Large language models (LLMs) offer the potential to automate a large
number of tasks that previously have not been possible to automate,
including some in science. There is considerable interest in whether
LLMs can automate the process of causal inference by providing the
information about causal links necessary to build a structural model. We
use the case of confounding in the Coronary Drug Project (CDP), for
which there are several studies listing expert-selected confounders that
can serve as a ground truth. LLMs exhibit mediocre performance in
identifying confounders in this setting, even though text about the
ground truth is in their training data. Variables that experts identify
as confounders are only slightly more likely to be labeled as
confounders by LLMs compared to variables that experts consider
non-confounders. Further, LLM judgment on confounder status is highly
inconsistent across models, prompts, and irrelevant concerns like
multiple-choice option ordering. LLMs do not yet have the ability to
automate the reporting of causal links.
\end{abstract}
\ifdefined\Shaded\renewenvironment{Shaded}{\begin{tcolorbox}[boxrule=0pt, sharp corners, frame hidden, interior hidden, borderline west={3pt}{0pt}{shadecolor}, enhanced, breakable]}{\end{tcolorbox}}\fi

\hypertarget{introduction}{%
\section{Introduction}\label{introduction}}

Generative artificial intelligence models, including large language
models (LLMs) like ChatGPT and its variants and competitors, offer
computers the ability to produce plausible text, images, and video. As
of this writing, there is enormous interest in finding areas in which
this new capability can be fruitfully applied or in which it should be
avoided. LLMs have already found applications in research and the
sciences generally (for just one example, Zhang et al. 2024). However,
there are reasons to doubt the value of LLMs in research making causal
claims using observational data. Properly performing causal inference
relies on theoretical understanding of real-world counterfactuals, while
LLMs are instead trained to replicate actual observed text, and it is
unclear the extent to which these models contain anything like a theory
capable of considering counterfactuals or latent causal factors in
real-world settings.

There is extensive work already exploring the causal reasoning capacity
of LLMs (e.g., Cai, Liu, and Song 2024; Kıcıman et al. 2023; Liu et al.
2024; Ashwani et al. 2024; Takayama et al. 2024; Jin et al. 2023; Han
2024; Sheth, Abdelnabi, and Fritz 2024). This line of research
investigates the ability of different LLM models to reason about a
causal link between variables, using whatever understanding it has of
the real world, or ability to mimic that understanding, in order to
determine the presence and direction of a causal link. This work either
tests performance of LLMs on these tasks, or investigates which other
automated tools, like causal discovery or knowledge databases, LLMs can
either aid in using or that LLMs need in order to effectively perform
causal reasoning. Success varies across models, approaches, and studies.

The aforementioned approaches rely on the ability of LLMs to reason
causally, which may or may not be an emergent property of LLMs but is
not an ability that their design ensures they will have. We take another
approach. LLMs are trained on large amounts of text data. Some of that
text data likely contains causal assertions. Even if LLMs would not be
able to reason their way to these assertions, they may be capable of
repeating them back, in effect acting as a large database of information
about causal links.

For example, LLMs do not need to be able to reason causally about the
effect of hormone replacement therapy on cardiovascular health to
determine that socioeconomic status (SES) is a confounder for this
relationship, since it has access to research studies (Humphrey, Chan,
and Sox 2002), and discussion of those studies (Catalogue of Bias
Collaboration et al. 2018), that explain SES is a confounder in this
context. So ``socioeconomic status'' becomes a statistically very likely
continuation of the text ``A good example of a confounder when studying
the effects of hormone replacement therapy on cardiovascular health
is'', and indeed this was the first continuation response we got when
giving this prompt to GPT-4o.

We aim to test whether LLMs can return causal judgements about a set of
potential confounders. Specifically, we examine the Coronary Drug
Project (CDP), an experimental study with imperfect compliance for which
expert opinion about the covariates necessary to adjust for imperfect
compliance are both available to us as a ground truth, and potentially
in the LLM's training data for them to report back. If the LLMs can
match expert opinion, then a non-expert researcher would be capable of
using the LLM responses to construct an expert-guided causal diagram and
use it to perform effective causal inference.

This is not the first study to take a recall-based approach to using
LLMs to make judgments about causal links. Long, Schuster, and Piché
(2023) identify a small set of known ground-truth causal links in the
medical literature and examine the ability of GPT-3 to match them. They
find some success, but a non-negligible error rate implying that results
would need to be further expert-verified, as well as heavy sensitivity
to prompt design. Zečević et al. (2023) refer to this theory of treating
LLMs as a causal link database as treating LLMs as ``causal parrots''
and find fairly poor performance from the GPT-3, Luminous, and OPT
models in matching ground-truth basic causal links, as well as variation
in results across models and prompt design, but that performance
improves somewhat from the injection of ground-truth facts into the
data, supporting a causal-link-recall theory.

We expand on this approach in several ways. First, we use a more complex
real-world setting in which we are attempting to build a single complex
model for performing causal inference, as a researcher would be likely
to do in application, rather than relying on ``toy'' causal links, on a
small set of ground truth links, or on a series of disconnected ground
truth links. Second, we use a more recent set of LLM models for which we
can establish that our intended ground-truth answers do exist in their
training data.

\hypertarget{the-coronary-drug-project}{%
\section{The Coronary Drug Project}\label{the-coronary-drug-project}}

The Coronary Drug Project (CDP) was a large placebo-controlled
randomized drug trial active between 1965 and 1985 investigating
treatments to reduce mortality among men with a history of myocardial
infarction (NHLBI BioLINCC 2024). In particular, we focus on the placebo
arm of the study. In the placebo arm, adherence to the assigned protocol
was imperfect, and there was considerable confounding between adherence
and mortality (CDPRG 1980).

In 1980, the degree of confounding was thought to be intractable: the
baseline difference in mortality rates across treatment adherence levels
in the placebo arm was 13.1\%, and inclusion of covariates shrank this
difference only to 9.4\% (CDPRG 1980). However, follow-up analyses were
able to use improved confounder selection and statistical adjustment
methods to reduce the difference to a more manageable 2.5\% (Murray and
Hernán 2016, 2018). Further studies have used causal discovery tools and
distributed expert opinion to further refine covariate selection
procedures (Gururaghavendran and Murray 2024; Debertin et al. 2024).

Because this case study has been examined for the specific case of
covariate selection for causal estimation multiple times, it serves as a
useful ground-truth case against which to compare LLM designations of
confounder status.

In addition to offering a ground-truth status, the prominent role of the
CDP means that studies concerning confounding in the CDP are likely to
be in the LLM training data. This is desirable for our case because we
are interested in whether LLMs can recall expert-claimed causal linkages
from its training data. If discussion of confounding in the CDP is in
the training data, then we know that discussions of expert opinions on
causal links for the variables relevant to the CDP are in the training
data.

As a basic test of whether discussions of confounding in the CDP are in
the LLM training data, we asked both GPT-4o and Claude about confounding
in the Coronary Drug Project, specifically ``What are the confounders
suggested to be necessary in adjusting for compliance in the Coronary
Drug Project?'' In their responses, both GPT-4o and Claude mentioned
that confounding was present in the placebo arm of the trial, which was
not given in the prompt. In their confounder suggestions, although
non-comprehensive and too generic in the case of this prompt to compare
directly, included many of those mentioned in CDPRG (1980). This implies
that in both models, at the very least CDPRG (1980) or commentary on
that paper are in the training data. Both models have at least some
expert opinion on all relevant causal links in their training data.

\hypertarget{methods}{%
\section{Methods}\label{methods}}

Our general approach to checking LLM capabilities to designate variables
as confounders involves taking a set of confounders identified by
experts, and for each having the LLM designate whether or not it is a
confounder.

We use confounder sets from three different studies. Confounders
included in the original CDPRG (1980) study are referred to as the
``Original'' confounder set. The re-analysis in Murray and Hernán (2016)
included a baseline/follow-up confounder distinction and re-evaluated
the set of confounders using a modern understanding of causal inference.
Variables included in Murray and Hernán (2016) but not CDPRG (1980) are
the ``Added in 2016'' confounder set. We also use the expert-guided
covariate set from Debertin et al. (2024), which used domain expertise
to add additional confounders to a ``maximal'' causal diagram, and then
produced a trimmed causal diagram after removing confounders that had
been considered but for which there was no evidence of a link, which we
refer to as ``Trimmed'' variables. Variables newly added in Debertin et
al. (2024) are ``Expert-Added.''\footnote{Since the focus is on the
  conceptual designation of confounder status, we do not distinguish
  between variables that were present in the original data set and those
  that were not present but still determined theoretically to be
  confounders.} Finally, we have a set of 60 variables that are in the
CDP dataset but are very unlikely to be confounders, which we call
``Non-confounders''. These variables were not selected by any of the
prior studies as likely confounders and fall into four main categories:
administrative variables; anticipated side-effects, or known
metabolites, of active CDP treatments; general medical information
collected from routine physical examination; and sub-study data
collected only on a sample of patients. See Appendix C for more
information.

We take several different approaches to designing prompts to query LLMs
about whether a given variable is a confounder in this context. First,
we distinguish between a ``direct'' prompting approach and an
``indirect'' prompting approach.

In the direct approach, we ask the LLM whether a given variable is a
confounder, in the following format:

\begin{quote}
I have a data set consisting only of people who have been assigned to
take a certain medication X. However, some of the patients choose to
take X as assigned, and others do not. I want to use causal analysis to
figure out whether (a patient's decision to take X) has a causal effect
on (the patient's mortality).

In this analysis, the variable (\{confounder\}) is\ldots{}

A. Not a confounding variable\\
B. A confounding variable\\
C. Not sure.
\end{quote}

Under this direct approach, a variable is determined by the LLM to be a
confounder if it responds with the option ``B.''

In the indirect approach, we generate two queries for each potential
confounder, in the first prompt asking about the relationship between
the confounder and adherence, and in the second asking about the
relationship between the confounder and mortality, in the following
format:

\begin{quote}
I have a data set consisting only of people who have been assigned to
take a certain medication X. However, some of the patients choose to
take X as assigned, and others do not. In this data, does
(\{confounder\}) have a causal effect on (mortality)?

A. Yes. (\{confounder\}) has a causal effect on (mortality).\\
B. No.~(\{confounder\}) and (mortality) are statistically related, but
there is not a causal relationship between them in either direction.\\
C. No.~(\{confounder\}) and (mortality) are not statistically or
causally related to each other.\\
D. Unsure.
\end{quote}

and similarly for adherence, with ``(a patient's decision to take X)''
in place of ``(mortality)''. In this indirect method, a given variable
is determined by the LLM to be a confounder if the LLM returns options
of either ``A.'' or ``B.'' for both the adherence version of the
question and the mortality version of the question.

We do not limit the prompt to only the question, however. Prior to
asking the question, we inform the LLM that it is an expert in the
domain of identifying confounders in medical studies, and provide three
additional example versions of the question the LLM will be answering,
along with correct answers. Following the question, we encourage the LLM
to engage in step-by-step reasoning and provide formatting instructions
for the response. Full prompts are shown in Appendix B.

Given our list of candidate confounders, we give these prompts to
generalist LLM models, specifically GPT-4o, GPT-o1-preview (which is
designed to engage in multi-step reasoning), and Claude 3.5 Sonnet. We
use generalist models, as opposed to models trained specifically on
medical data like MedLlama3 (see Zhou et al. 2023), for several reasons.
First, medical LLMs, while they do contain expert medical knowledge, are
generally trained for medical practitioner tasks like diagnostics rather
than observational or behavioral research, which for an adherence study
is more relevant. Second, our goal is to test the causal-recall
capabilities of LLMs generally, and many relevant areas of application
will not have specialized LLMs available. Third, as discussed in the
previous section, we have reason to believe that the relevant domain
knowledge is in the generalist LLM data sets. Fourth, investigation by
Nori et al. (2023) finds that the current cohort of generalist models,
when encouraged to reason step-by-step, are capable of exhibiting
medical domain knowledge of similar quality to the specialized LLM
models.\footnote{An earlier version of this paper, based on an earlier
  cohort of LLMs, included the science-specialist Galactica and the
  biomedical-specialist BioMedLM models. In that version we did not find
  that these specialized models outperformed the generalist GPT-3,
  3.5-turbo, or 4 models.}

GPT-4o and Claude 3.5 Sonnet are run using a temperature of 0.7, and
each prompt is given ten times, in order to allow us to see some of the
distribution of responses that the LLMs might give.\footnote{In the case
  of the indirect prompts, each response relating to adherence is
  matched with each response relating to mortality one at a time. The
  confounder designation is determined for each pair of responses, and
  averaged to give the share of the time that the LLM designates the
  variable as a confounder.} GPT-o1-preview does not allow users to
raise the temperature. These responses are then compared to the
expert-chosen confounder lists from the literature.

The prompting method is slightly different for the ``Non-confounder''
group for GPT-o1-preview queries. The ``Non-confounder'' group was the
last set of variables queried, and was first queried as though all
variables were measured-at-baseline before some were corrected to
measured-at-followup. Before the measured-at-followup correction was
made, the GPT-o1-preview content flagging system updated, and our
prompts were consistently flagged as inappropriate. We were able to get
responses to prompts using the Direct prompting method by adding ``I am
asking this question purely for the purposes of performing statistical
research. I am NOT asking for medical advice'' to the end of the prompt,
but were unable to find a way to adjust the Indirect-method prompt that
would avoid content flagging. As such, for GPT-o1-preview,
``Non-confounders'' were queried using this additional line in the
prompt, and also Non-confounders measured at a non-baseline time were
instead prompted as though the variable was measured at baseline (see
Appendix A). This only affects Non-confounders, does not affect any
GPT-4o or Claude Sonnet 3.5 output, and in the case of Direct-method
prompting only slightly changes the prompt.

In order to assess the consistency of LLM responses we also use several
variants on the prompts. First, we attempt a ``no-reasoning'' version of
the prompts, in which the LLMs are told they are an expert in
identifying confounders in medical studies and then given the question
without any example tasks. Then the LLM is instructed to give its
response immediately rather than providing any explanation or reasoning.

Second, we modify the direct prompt and change the order of the options
to ``A. Not sure. B. Not a confounding variable. C. A confounding
variable'' to check whether the LLM responses are sensitive to the
shuffling of the options, as in Nori et al. (2023).

One alternative approach we do not take is to design the prompt to
mention the CDP. We could modify the prompt to specify that we want to
know whether a given variable was found to be a confounder in the CDP,
or even in CDPRG (1980) specifically. We already did this to some extent
in the previous section when evaluating whether the CDP was in the LLM
training data. This might hew more closely to investigating LLM ability
to recall specific facts from research papers. However, we are
interested in the LLMs' ability to recall text containing causal claims
more broadly, and in most applications there are not papers specifying
confounder lists for a specific intervention. We want the LLM to pull
from text concerning confounder selection \emph{like} the papers we use
as ground truth or other text discussing confounding and causality. But
for this to be usable generally, the LLM must be able to use context,
rather than a specific paper or intervention name, to determine on its
own what text it should be looking at. For similar reasons, we do not
give the LLMs the text of the three studies as a part of our prompt.

\hypertarget{data}{%
\section{Data}\label{data}}

From CDPRG (1980), Murray and Hernán (2016), and Debertin et al. (2024)
we have a list of 172 potential confounders to consider across eight
different prompt variations. For the Claude 3.5-Sonnet and GPT-4o
models, we query the LLM ten times each for a total of 13,760
LLM-generated responses each. For the GPT-o1-preview model, we only
query the LLM once, and only use four of the prompt variations (omitting
the no-reasoning prompts since GPT-o1-preview always provides reasoning)
for a total of 688 LLM-generated responses.

Responses for all confounder categories other than ``Non-Confounders''
were gathered between October 9 and 15, 2024 using the Python packages
\texttt{openai} (OpenAI 2024) and \texttt{anthropic} (Anthropic 2024).
``Non-confounder'' responses were collected between November 23 and
December 8, 2024.

Table \ref{tbl-descriptives} shows the distribution of how all the
confounders were classified across the range of prompts, averaging over
all ten iterations for the Claude 3.5-Sonnet and GPT-4o models. We see
that, in general, the LLMs designated variables as confounders most of
the time, especially when the indirect method was used. Roughly 90\% of
confounders were designated as confounders by Claude Sonnet-3.5 and by
GPT-o1-preview using the indirect method, with 81.9\% designated as
confounders with GPT-4o using the indirect method. 58-72\% were
designated as confounders using the direct method across all three
models.

\hypertarget{tbl-descriptives}{}
\begingroup
\fontsize{12.0pt}{14.4pt}\selectfont
\begin{longtable}{>{\raggedright\arraybackslash}p{\dimexpr 0.20\linewidth -2\tabcolsep-1.5\arrayrulewidth}>{\raggedleft\arraybackslash}p{\dimexpr 0.11\linewidth -2\tabcolsep-1.5\arrayrulewidth}>{\raggedleft\arraybackslash}p{\dimexpr 0.11\linewidth -2\tabcolsep-1.5\arrayrulewidth}>{\raggedleft\arraybackslash}p{\dimexpr 0.11\linewidth -2\tabcolsep-1.5\arrayrulewidth}>{\raggedleft\arraybackslash}p{\dimexpr 0.11\linewidth -2\tabcolsep-1.5\arrayrulewidth}>{\raggedleft\arraybackslash}p{\dimexpr 0.11\linewidth -2\tabcolsep-1.5\arrayrulewidth}>{\raggedleft\arraybackslash}p{\dimexpr 0.11\linewidth -2\tabcolsep-1.5\arrayrulewidth}>{\raggedleft\arraybackslash}p{\dimexpr 0.11\linewidth -2\tabcolsep-1.5\arrayrulewidth}}
\caption{\label{tbl-descriptives}Share of Variables Designated as Confounders }\tabularnewline

\toprule
Model & Is Confounder (Dir.) & Unsure & Is Confounder (Indir.) & Related to Treatment & Unsure Rel. Treat. & Related to Outcome & Unsure Rel. Outcome \\ 
\midrule\addlinespace[2.5pt]
Claude 3.5 & 59.9\% & 13.8\% & 89.5\% & 91.0\% & 3.1\% & 93.7\% & 1.7\% \\ 
GPT-4o & 71.9\% & 10.2\% & 81.9\% & 86.6\% & 7.7\% & 93.2\% & 4.6\% \\ 
GPT-o1 & 58.7\% & 0.0\% & 93.6\% & 93.6\% & 2.3\% & 100.0\% & 0.0\% \\ 
\bottomrule
\end{longtable}
\endgroup

\hypertarget{results}{%
\section{Results}\label{results}}

\hypertarget{confounder-designation-with-reasoning}{%
\subsection{Confounder Designation with
Reasoning}\label{confounder-designation-with-reasoning}}

\begin{figure}

{\centering \includegraphics{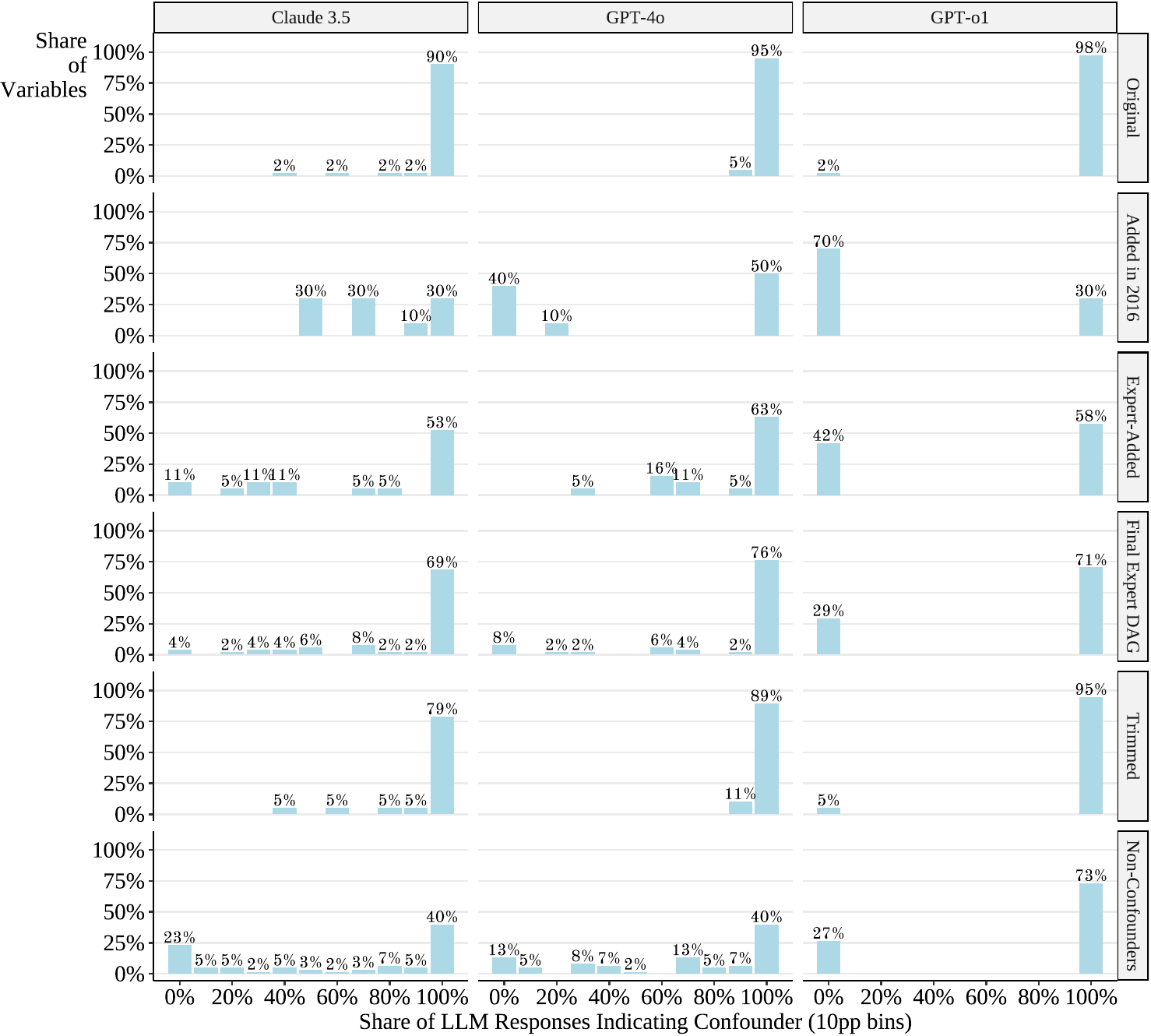}

}

\caption{\label{fig-agree-with-expert-direct}Distribution of LLM
Confounder Designations with Reasoning-Encouraged Direct Method}

\end{figure}

\begin{figure}

{\centering \includegraphics{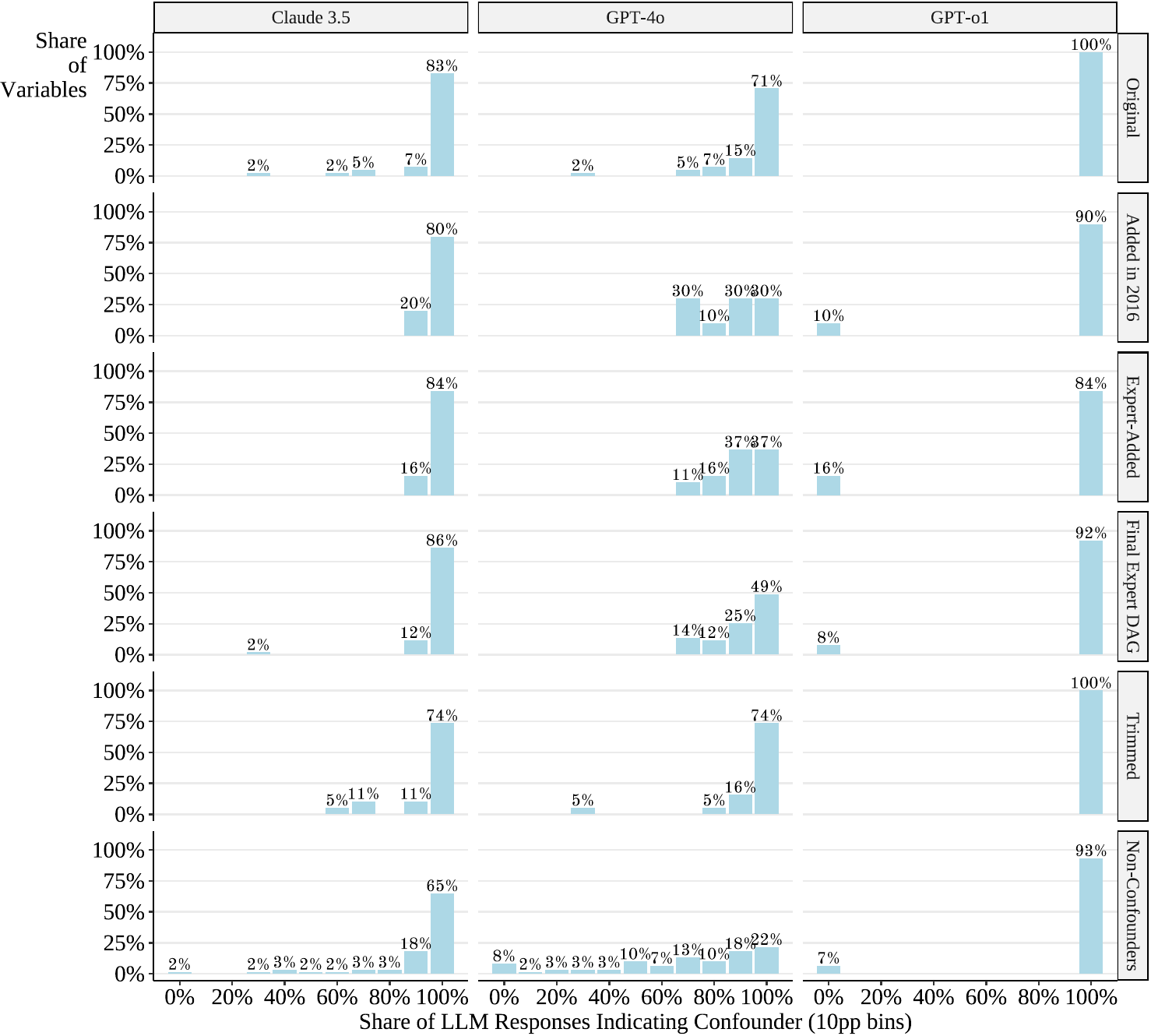}

}

\caption{\label{fig-agree-with-expert-indirect}Distribution of LLM
Confounder Designations with Reasoning-Encouraged Indirect Method}

\end{figure}

The distributions of LLM confounder designations by model and confounder
set are shown in Figure~\ref{fig-agree-with-expert-direct} for
designations made using the direct method (where the LLM is directly
asked whether the variable is a confounder), and in
Figure~\ref{fig-agree-with-expert-indirect} for designations made using
the indirect method (where the LLM is separately asked about the
relationship between the potential confounder and adherence, and between
the potential confounder and mortality).

Figure~\ref{fig-agree-with-expert-direct} shows strong ability to
designate the confounders identified in CDPRG (1980) as confounders
(``Original''). Beyond this, the LLM's ability to match expert opinion
is considerably weaker. LLMs do not generally classify the additional
confounders added to CDPRG (1980) by Murray and Hernán (2016) as
confounders, with only GPT-4o including even 50\% of them (``Added in
2016''). Perhaps oddly, the confounders trimmed from the causal diagram
in Debertin et al. (2024) as being unnecessary were far more likely to
be classified by LLMs as confounders than the variables added and kept.
The variables directly classified as Non-Confounders were also more
likely to be designated as confounders than the Added in 2016 set,
except for GPT-4o. That variables removed from the causal diagram were
more likely to be designated confounders than the variables kept, and
that variables designated as non-confounders were included at a similar
rate, indicates disagreement between the LLMs and the experts

There is more agreement between the LLMs and the final set of variables
in Debertin et al. (2024), but even in this case only about 70\% of
confounders in their final causal diagram were designated by LLMs as
confounders. The rates are lower - 53-63\% - for confounders directly
added by the experts.

Overall, the direct method of eliciting confounder designations led to
fairly high degrees of agreement with CDPRG (1980), but unimpressive
degrees of agreement with either of the more modern follow-up studies,
which use a more up-to-date understanding of causal inference and
variable selection.

Figure~\ref{fig-agree-with-expert-indirect} shows confounder
designations made using the indirect method.
Figure~\ref{fig-agree-with-expert-indirect} shows considerable variation
across models, with both Claude 3.5-Sonnet and GPT o1-preview
designating very high shares of variables as confounders across all
variable sets. GPT-4o, on the other hand, is more reluctant to designate
variables as confounders.

Claude 3.5-Sonnet and GPT o1-preview both identify a high share of
variables as confounders. This indicates a high degree of agreement with
the confounders identified in all three studies. However, the rate of
confounder designation is nearly as high for the ``Trimmed'' group of
variables removed from the causal diagram as unnecessary and the
``Non-Confounder'' group considered by experts not to be confounders. So
this pattern seems to indicate less a strong agreement with the original
studies, and more that the indirect method in these two models produces
a high likelihood of returning a positive designation.

GPT-4o with an indirect prompt is much less likely to designate a
variable as a confounder. Oddly, it designates the highest share of
variables as confounders in the Trimmed group of variables, where, if
GPT-4o were agreeing with the experts, we would expect a fairly low rate
of agreement. GPT-4o does designate a smaller share of Non-Confounders
as being confounders than in most other categories, but this is
inconsistent: a small share of Non-confounders are identified as
confounders 100\% of the time the prompt is given (the far-right bar in
the graph), but in many cases the same prompt produces inconsistent
results: for more than half of Non-confounders, the variable is
designated as a confounders more than 50\% the time the prompt is given.

\hypertarget{confounder-designation-without-reasoning}{%
\subsection{Confounder Designation Without
Reasoning}\label{confounder-designation-without-reasoning}}

\begin{figure}

{\centering \includegraphics{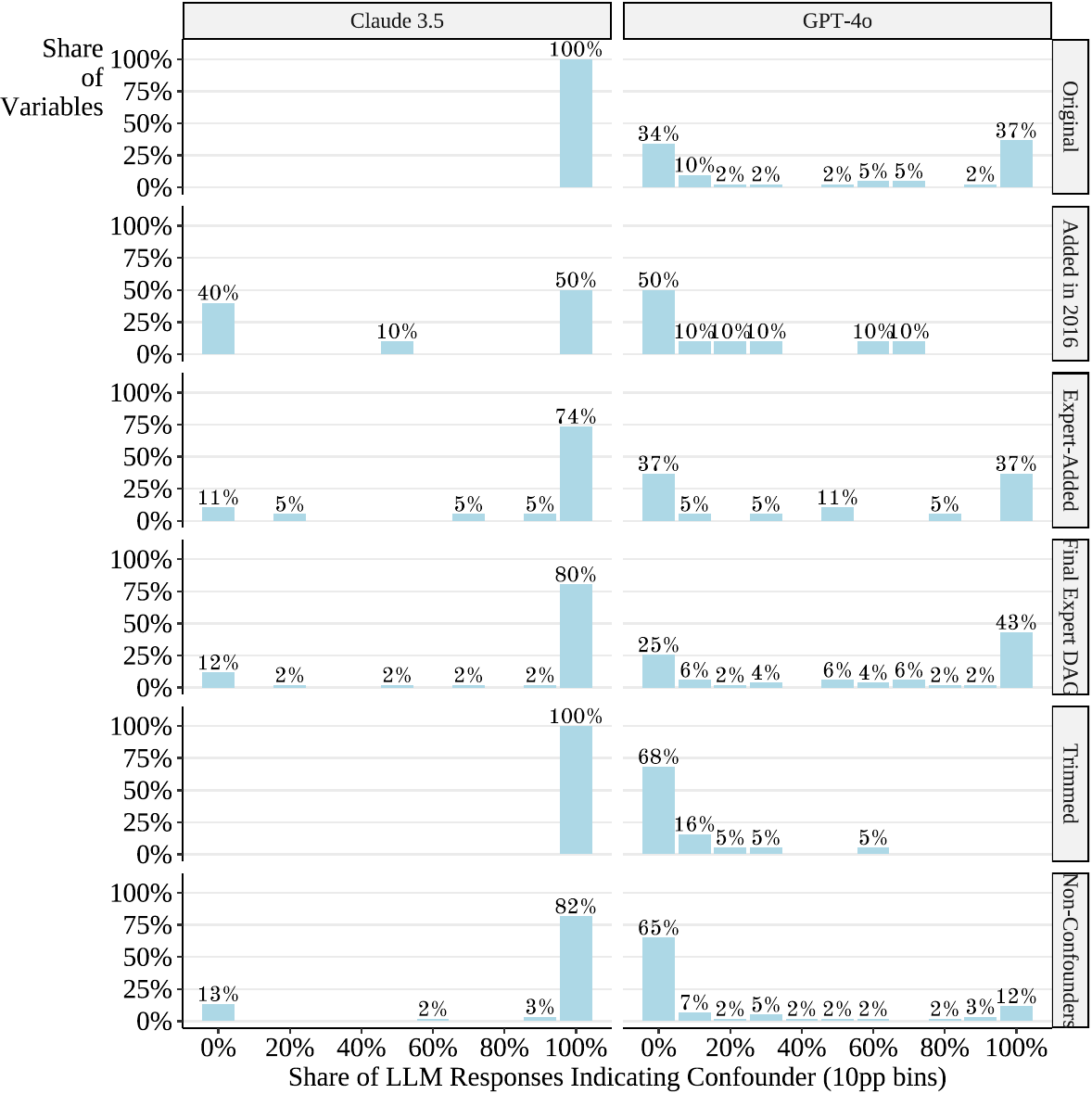}

}

\caption{\label{fig-agree-with-expert-direct-noreas}Distribution of LLM
Confounder Designations with No-Reasoning Direct Method}

\end{figure}

\begin{figure}

{\centering \includegraphics{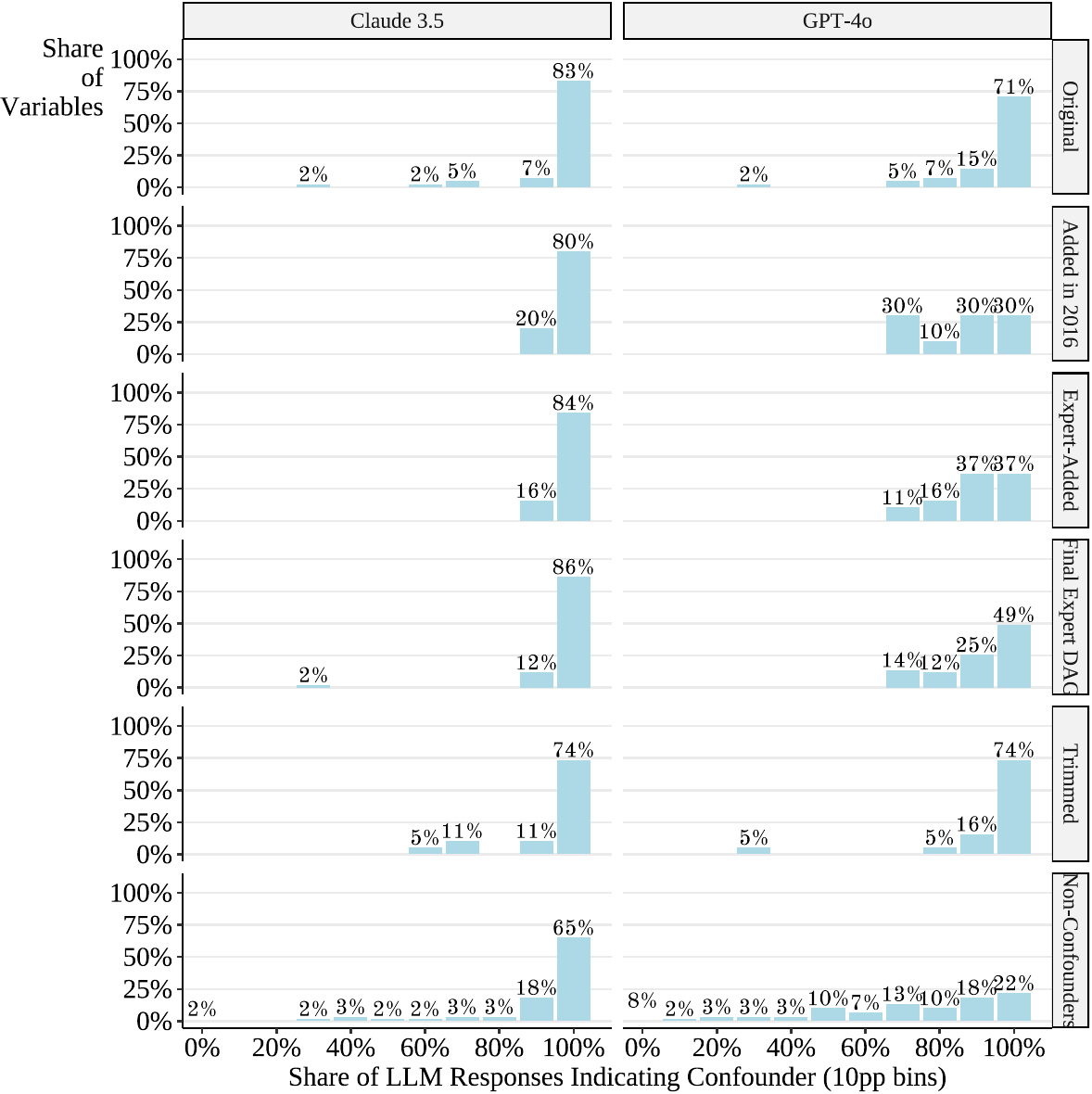}

}

\caption{\label{fig-agree-with-expert-indirect-noreas}Distribution of
LLM Confounder Designations with No-Reasoning Indirect Method}

\end{figure}

Figure~\ref{fig-agree-with-expert-direct-noreas} and
Figure~\ref{fig-agree-with-expert-indirect-noreas} repeat
Figure~\ref{fig-agree-with-expert-direct} and
Figure~\ref{fig-agree-with-expert-indirect} but with prompts that
discourage the LLM from walking through its logic step-by-step and
instead instruct it to produce an answer immediately.

Responses in Figure~\ref{fig-agree-with-expert-direct-noreas} are
somewhat erratic. Claude 3.5-Sonnet identifies 100\% of the confounders
from CDPRG (1980), but only half of those kept in Murray and Hernán
(2016), all of the Trimmed variables, and nearly all of the
Non-Confounders.

GPT-4o shows a wide range of probabilities, with a great number of
variables being designated as confounders in some iterations and not in
others. It never identifies more than half of the variables in any group
as confounders, although it also generally rejects the Trimmed variables
and the Non-Confounders. This may just be due to its reduced tendency to
designate a variable as a confounder at all, given its high
false-negative rate.

Figure~\ref{fig-agree-with-expert-indirect-noreas} shows very similar
patterns to Figure~\ref{fig-agree-with-expert-indirect}. For Claude
3.5-Sonnet, a high share of variables are designated as confounders,
including those in the Trimmed and Non-Confounders group, although these
are designated at a slightly lower rate than under the direct method.
GPT-4o finds the highest share of confounders in the Trimmed group.
After that, it designates 71\% of the CDPRG (1980) confounder set as
confounders. It only designates 22\% of the Non-Confounders as
confounders 100\% of the time, although there is little consistency
across iterations, and it designates 80\% of Non-Confounders as
confounders at least half the time.

\hypertarget{consistency-of-llm-confounder-designations}{%
\subsection{Consistency of LLM Confounder
Designations}\label{consistency-of-llm-confounder-designations}}

In this section, instead of comparing LLM confounder designations to the
expert-determined ground truth, we consider how robust the confounder
designations are to repeated prompting or slight variations in
prompting.

\begin{figure}

{\centering \includegraphics{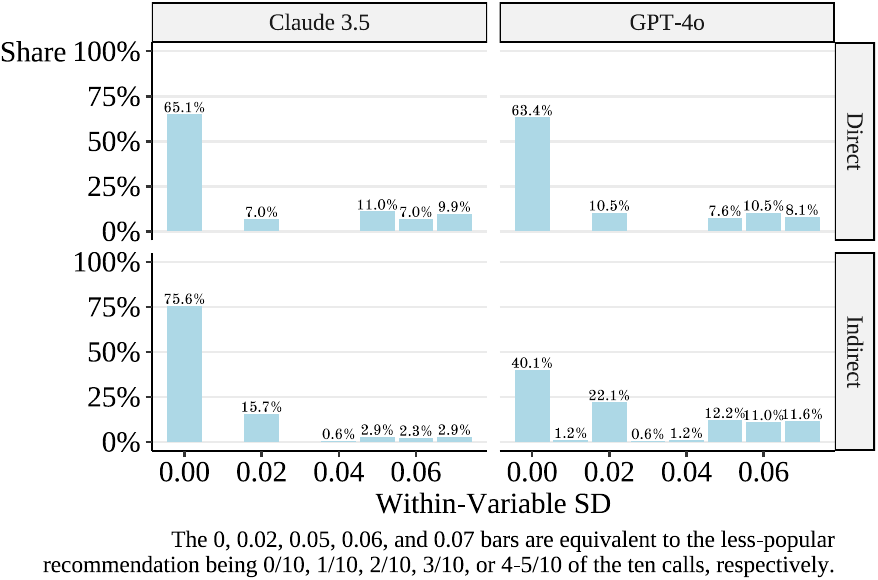}

}

\caption{\label{fig-iterations}Consistency of Confounder Designations
Across Iterations}

\end{figure}

For Claude 3.5-Sonnet and GPT-4o, each prompt was given 10 times, using
a temperature setting of .7. Figure~\ref{fig-iterations} shows the
consistency of LLM designations across iterations of the same
with-reasoning prompt.\footnote{Keep in mind that the reasoning prompts
  have the LLM produce a detailed explanation of its thinking before
  selecting an answer. Each different prompt generation produces a
  different explanation, and the explanation it happens to generate will
  lead to a different distribution of multiple-choice responses. So it
  is not feasible to simply show the internal probability of each
  multiple-choice token and take that as a distribution of response
  probabilities. We must actually generate the responses from the start.}
Consistency across iterations varies. Claude reported the same result
for all ten iterations for 65.1\% of variables under the direct method,
and for 75.6\% of all variables under the indirect method.\footnote{Notably,
  for both Claude and GPT, there was more inconsistency in designation
  among the Non-Confounder group. If this group is excluded, the shares
  for Claude are 66.1\% direct and 81.2\% indirect, and for GPT-4o are
  68.8\% direct and 48.2\% indirect.} GPT-4o was more sensitive to
method. 63.4\% of all variables produced the same result in every
iteration under the direct method, but this fell to just 40.1\% for the
indirect method, with 22.6\% of the sample splitting the ten iterations
4-6 or 5-5 in their designation. Over these iterations we see modest
consistency across iterations, although there is not a strong pattern as
to which model or method produced the most consistency, and a
non-negligible portion of the sample produced a split conclusion.

The rest of the results in this section look at whether, for a given
variable, two different approaches to prompting produce the same result.
For each variable and prompting method (or model), the variable is
categorized based on whether it is never (0\%), always (100\%), or
sometimes (Mixed) designated as a confounder. We then look at the share
of variables that, for example, are 100\% confounders under one method
but Mixed under another method. Non-confounders measured at follow-up
are omitted from this graph for the GPT-o1-preview model, for the data
inconsistency reasons outlined in the Data section.

\begin{figure}

{\centering \includegraphics{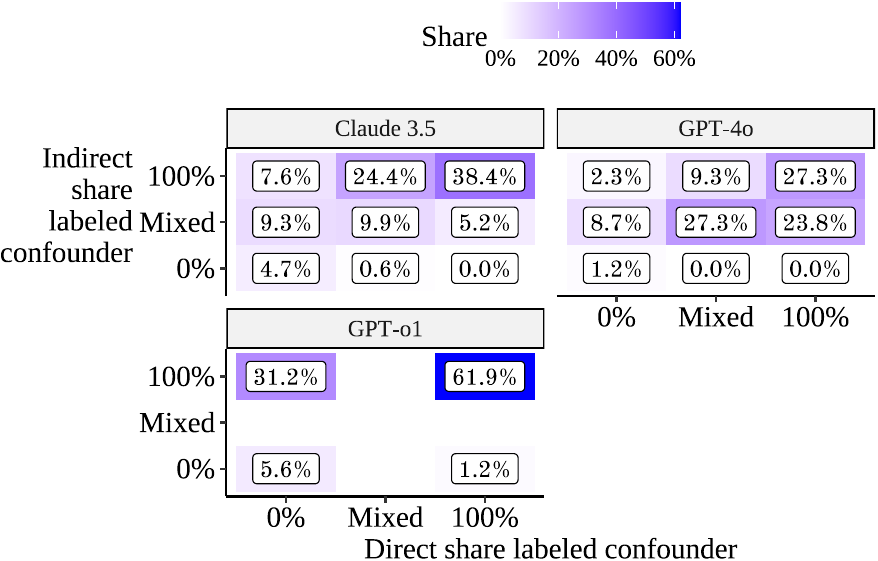}

}

\caption{\label{fig-direct-vs-indirect}Confounder Designation Agreement
Between Direct and Indirect Prompting Methods With Reasoning}

\end{figure}

In the previous section we already found that the distribution of
confounder designations differed between direct and indirect prompting.
Figure~\ref{fig-direct-vs-indirect} shows the distribution of agreement
by variable.\footnote{Chi-square tests of a statistically significant
  relationship between the two prompting methods are not included here
  since these tests are not identified when one of the rows or columns
  contains only zero values, which happens frequently in this case.} For
GPT o1-preview, there was fair agreement across methods, with 87.7\% of
the variables designated the same way; however, this was driven
partially by the model's tendency to label everything a confounder, so
the Cohen's kappa was still low at .16. Claude 3.5 Sonnet saw only
66.9\% of confounders getting the same designation in both methods, and
Cohen's kappa of .21. In both, the mismatches usually occurred because
the indirect method designated something as a confounder while the
direct method did not. GPT-4o produced a similar Cohen's kappa of .24,
although much of this was due to counting different non-0/non-100\%
designations as matches. Mismatch was in the opposite direction, usually
driven by the direct method designating something as a confounder when
the indirect method did not.

\begin{figure}

{\centering \includegraphics{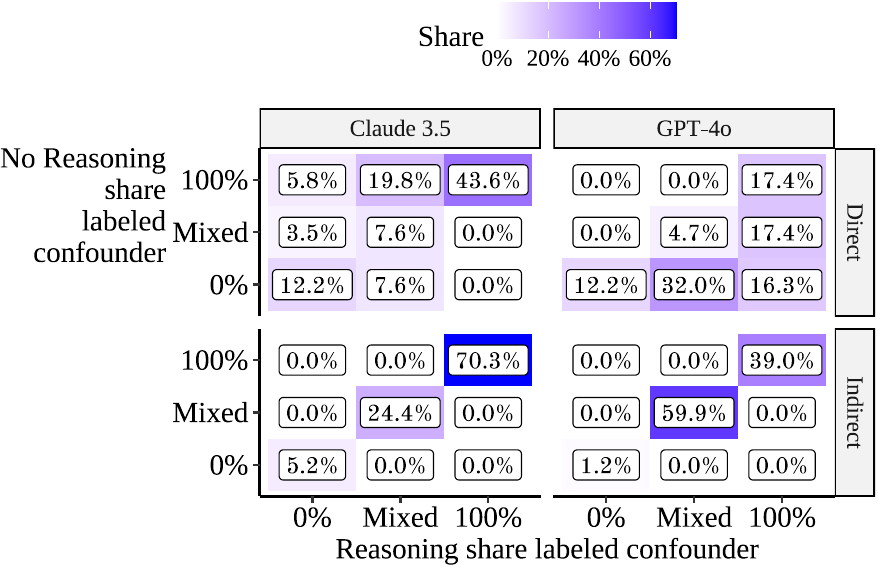}

}

\caption{\label{fig-reas-vs-non}Confounder Designation Agreement Between
Reasoning and No-Reasoning Prompts}

\end{figure}

Figure~\ref{fig-reas-vs-non} compares designations made by the reasoning
and no-reasoning versions of the prompts. The degree of agreement across
reasoning vs.~no-reasoning is highly sensitive to the interaction with
direct vs.~indirect prompting. Reasoning vs.~no-reasoning made no
difference at all for indirect prompting, producing Cohen's kappa values
of 1.\footnote{Agreement is not actually perfect, since the ``Mixed''
  designation includes the entire range of percentages between 0\% and
  100\% (non-inclusive) and so a 10\% in one method but 90\% in the
  other would be regarded as a match. But agreement is perfect for the
  specified bins.} On the other hand, there was little agreement for
direct prompting in GPT-4o, with a Cohen's kappa value of .13. Agreement
was higher for Claude Sonnet 3.5, with a Cohen's Kappa of .41. The
direction of mismatch varies: mismatches are more often due to the
reasoning method designating more confounders than the no-reasoning
method for GPT-4o, but the opposite is true for Claude 3.5 Sonnet.

The Figure~\ref{fig-reas-vs-non} result can be taken as mixed evidence
for our theory of how LLMs could designate confounders, although only
weakly. One might expect that if the LLMs are performing pure recall
from their text data, then results should be identical between reasoning
and no-reasoning prompts (as we find for indirect prompting). However,
if we take that claim seriously, we only find mixed evidence of it as it
does not replicate for direct prompting. That said, even if the LLM is
performing pure recall, we still might expect responses with reasoning
to differ, since LLMs generate token probabilities based on the prompt
text up to that point, including its own output. Allowing the LLM to
discuss the problem, even if it is not actually performing reasoning,
may push it into the part of its training data where the information
actually lies and improve its performance even if it is doing pure
recall. So either way, this is only weak evidence for or against our
proposed mechanism.

\begin{figure}

{\centering \includegraphics{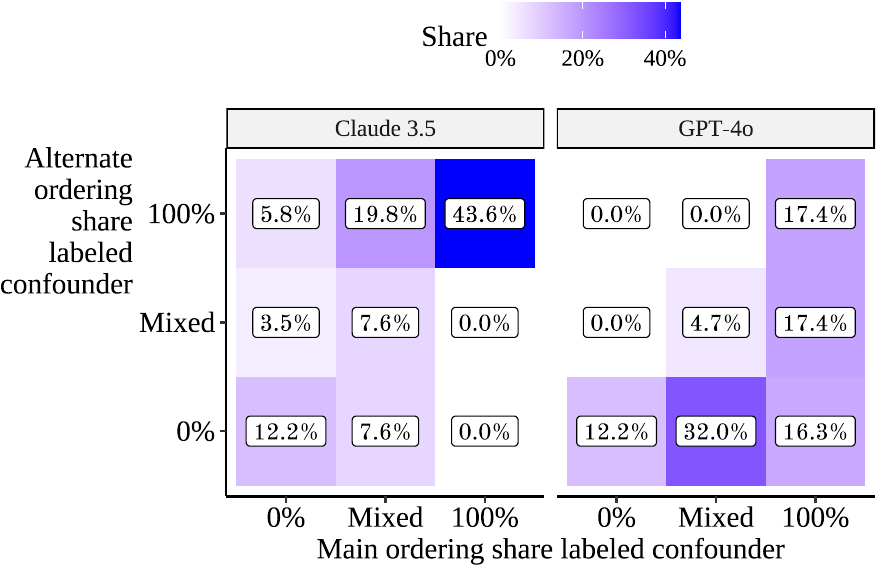}

}

\caption{\label{fig-option-order}Confounder Designation Sensitivity to
Option Ordering}

\end{figure}

For the direct version of the prompt with reasoning, we elicited a
second set of responses where the only change to the prompt was a change
in the order that the multiple-choice options were presented. This
allows us to test for order-option sensitivity, as in Nori et al.
(2023). Figure~\ref{fig-option-order} shows the results. We find
significant amounts of order-option sensitivity. In Claude 3.5, 36.7\%
of variables had different designations under one option-ordering system
compared to another, leading to a Cohen's kappa of .41. GPT-4o was less
consistent, with 65.7\% of variables changing designation, leading to a
Cohen's kappa of .13. These deviations based on option ordering were not
minor either, with 4.6\% of variables in Claude, and 16.3\% in GPA-4o,
switching between 0\% confounder designation and 100\% based only on
option ordering. These results suggest a heavy impact of option
ordering, which should not rationally affect the response that the LLM
gives.

\hypertarget{conclusion}{%
\section{Conclusion}\label{conclusion}}

In this paper, we use three mass-market LLMs to designate a set of
variables as confounders or not, using several different prompting
strategies. We use the Coronary Drug Project (CDP) as a context in which
to ask these questions, which allows us to use expert confounder
designations in CDPRG (1980), Murray and Hernán (2016), and Debertin et
al. (2024) as ground truths for comparison. Our working theory of how
the LLM would be able to designate variables as confounders relies on
the LLM having text data discussing confounder selection in its training
data (as we know it does for CDPRG (1980)), as opposed to any reliance
on the LLM being able to use real-world understanding to designate a
variable as a confounder.

We find some moderate but mixed success in having LLMs produce
confounder designations that match expert opinion. There are some cases
in which the LLMs are able to match expert opinion a large share of the
time, and in most cases, the LLM is able to categorize more than 70\% of
expert-designated confounders as confounders. However, the LLM also
designated as confounders a high number of variables in the ``Trimmed''
or ``Non-Confounders'' categories, which were, respectively removed from
the causal diagram as unnecessary, or specifically designated as
non-confounders. We would expect that these categories would be
designated as confounders at a lower rate, but they are often designated
at the same rate, and in some cases at higher rates. Success at matching
expert confounder designations seems to rely more on whether a given
model and prompting method tended to produce mostly-positive results or
mostly-negative results, without a heavy signal to distinguish
true-positives from true-negatives.

If we take as a rule of thumb a 70\%+ rate of being able to identify
variables as confounders (regardless of ability to identify
non-confounders as non-confounders), this level of success may be in
some cases acceptable. Further, it is possible that the LLMs are in fact
not recalling expert opinion from its text data and instead are doing a
\emph{better} job of applying causal reasoning to identify confounders
than the experts; we cannot prove that in the case of a deviation the
experts are more correct than the LLMs, only that they are not the same.

However, both of these potential justifications for LLM performance are
blunted by the considerable amounts of inconsistency we found in the LLM
designations, which reduce the practical usefulness of a 70\%+ match and
make it less likely that the LLM is reporting some truth about
confounders unknown to experts. Responses were sensitive to whether
prompts were direct (asking directly about confounding status)
vs.~indirect (asking separately about the relationship of the variable
with treatment and control). Multiple iterations of the same prompt
produced the same result as little as 40\% of the time in some cases,
although as much as 76\% in other cases. The quality of the output
varied across which LLM model was used, and were highly sensitive to
irrelevant details like what order the multiple-choice options for
confounder designation were presented in. In all of these cases, the
inconsistency sometimes favored one approach and other times favored its
opposite, so this inconsistency does not simply let us conclude that
we've discovered that one approach performs better than another.

These results show that available market LLMs achieve a non-negligible
success rate in identifying that a given variable is a confounder.
However, that success rate is also not high enough that it could be
meaningfully used to select confounders in a real-world setting; even in
some of our reasonably well-performing output, like Claude 3.5 Sonnet
with indirect prompting, expert-selected covariates had an 80-86\%
chance of being correctly designated as confounders, but expert-rejected
covariates had an 65-74\% chance of being incorrectly designated as
confounders. There is a signal here, but it is not strong enough that
one could confidently generate a list of confounders for an actual study
in this way, even if we knew ahead of time that Claude 3.5 Sonnet with
indirect prompting was a desirable set of methodological choices.

Additionally, results are sensitive to seemingly arbitrary changes in
prompting. Sensitivity to prompting can be acceptable in use cases where
a user could distinguish which prompt is producing better output, but
determining which output is better in the confounder-selection case
would require that experts simply do the job from top to bottom anyway,
making the LLM's contributions redundant. Further, even this level of
success occurs in the CDP context where the LLMs have been handed the
answer key ahead of time. One could reasonably believe that LLM
confounder selection in a less well-studied case would be less
effective.

These results of course cannot prove that LLMs can never complete this
task. LLM technology is constantly improving, and none of the results in
this paper prove that there will never be an LLM or a prompt design
capable of effectively selecting covariates. But this paper shows that
the technology does not as of yet have this capability, and the
inconsistency in results suggests that even if we did design a prompt
capable of matching experts in CDP in a certain model we should be
skeptical that that same prompt and model would work in other contexts.
We do not have to leave this to speculation, however: the setting and
ground-truth in this study provides a test that can be re-performed with
future models to determine whether success rates have indeed improved.
To that end, our code, which can be easily revised to test future LLM
tools, will be available at \url{https://osf.io/spzbu/}.

\hypertarget{appendix-a-full-variable-sets}{%
\section*{Appendix A: Full Variable
Sets}\label{appendix-a-full-variable-sets}}
\addcontentsline{toc}{section}{Appendix A: Full Variable Sets}

\hypertarget{original-confounder-set}{%
\subsection{Original Confounder Set}\label{original-confounder-set}}

Confounders measured at baseline: Age at entry, Antiarrhythmic agents,
Antihypertensives other than diuretics, Digitalis, Diuretics, History of
acute coronary insufficiency, History of angina pectoris, History of
congestive heart failure, History of intermittent cerebral ischemic
attack, History of intermittent claudication, Number of myocardial
infarctions at baseline, Oral hypoglycemic agents, Pre-randomization
adherence, Race, Relative body weight, Risk group, Time since most
recent myocardial infarction

Confounders measured at follow-up: Absolute neutrophil count,
Cardiomegaly on chest X-ray, Cigarette smoking, Current habitual level
of physical activity, Diastolic blood pressure, Heart rate, Hematocrit,
New York Heart Association functional class, Plasma fasting glucose,
Plasma one-hour glucose, after 75g oral load, Plasma urea nitrogen,
Premature ventricular beats, Q/QS pattern on antero-lateral,
postero-inferior, or antero-septal recording, Serum alkaline
phosphatase, Serum total bilirubin, Serum total cholesterol, Serum
triglyceride, Serum uric acid, ST segment depression on antero-lateral,
postero-inferior, or antero-septal recording, ST segment elevation on
antero-lateral, postero-inferior, or antero-septal recording, Systolic
blood pressure, T-wave findings on antero-lateral, postero-inferior, or
antero-septal recording, Ventricular conduction defect, White blood cell
count

\hypertarget{follow-up-confounder-set}{%
\subsection{2016 Follow-up Confounder
Set}\label{follow-up-confounder-set}}

Confounders measured at baseline: Age at entry, Antiarrhythmic agents,
Antihypertensives other than diuretics, Digitalis, Diuretics, History of
acute coronary insufficiency, History of angina pectoris, History of
congestive heart failure, History of intermittent cerebral ischemic
attack, History of intermittent claudication, Number of myocardial
infarctions at baseline, Oral hypoglycemic agents, Pre-randomization
adherence, Race, Relative body weight, Risk group, Time since most
recent myocardial infarction

Counfounders measured at follow-up: Absolute neutrophil count,
Antiarrhythmic agents, Antihypertensives other than diuretics,
Cardiomegaly on chest X-ray, Cigarette smoking, Current habitual level
of physical activity, Diastolic blood pressure, Digitalis, Diuretics,
Heart rate, Hematocrit, History of acute coronary insufficiency, History
of angina pectoris, History of congestive heart failure, History of
intermittent cerebral ischemic attack, History of intermittent
claudication, New York Heart Association functional class, Oral
hypoglycemic agents, Plasma fasting glucose, Plasma one-hour glucose,
after 75g oral load, Plasma urea nitrogen, Premature ventricular beats,
Q/QS pattern on antero-lateral, postero-inferior, or antero-septal
recording, Serum alkaline phosphatase, Serum total bilirubin, Serum
total cholesterol, Serum triglyceride, Serum uric acid, ST segment
depression on antero-lateral, postero-inferior, or antero-septal
recording, ST segment elevation on antero-lateral, postero-inferior, or
antero-septal recording, Systolic blood pressure, T-wave findings on
antero-lateral, postero-inferior, or antero-septal recording,
Ventricular conduction defect, White blood cell count

\hypertarget{confounder-set-after-removing-non-relevant-nodes}{%
\subsection{Confounder Set After Removing Non-Relevant
Nodes}\label{confounder-set-after-removing-non-relevant-nodes}}

Confounders measured at baseline: Age at entry, Antiarrhythmic agents,
Antihypertensives other than diuretics, Digitalis, Diuretics, History of
acute coronary insufficiency, History of angina pectoris, History of
congestive heart failure, History of intermittent cerebral ischemic
attack, History of intermittent claudication, Number of myocardial
infarctions at baseline, Oral hypoglycemic agents, Pre-randomization
adherence, Race, Relative body weight, Risk group, Time since most
recent myocardial infarction

Counfounders measured at follow-up: Antiarrhythmic agents,
Antihypertensives other than diuretics, Cardiomegaly on chest X-ray,
Cigarette smoking, Current habitual level of physical activity,
Digitalis, Diuretics, History of acute coronary insufficiency, History
of angina pectoris, History of congestive heart failure, History of
intermittent cerebral ischemic attack, History of intermittent
claudication, New York Heart Association functional class, Oral
hypoglycemic agents, Plasma fasting glucose

\hypertarget{trimmed-variables}{%
\subsection{Trimmed Variables}\label{trimmed-variables}}

Confounders measured at follow-up: Absolute neutrophil count, Diastolic
blood pressure, Heart rate, Hematocrit, Plasma one-hour glucose, after
75g oral load, Plasma urea nitrogen, Premature ventricular beats, Q/QS
pattern on antero-lateral, postero-inferior, or antero-septal recording,
Serum alkaline phosphatase, Serum total bilirubin, Serum total
cholesterol, Serum triglyceride, Serum uric acid, ST segment depression
on antero-lateral, postero-inferior, or antero-septal recording, ST
segment elevation on antero-lateral, postero-inferior, or antero-septal
recording, Systolic blood pressure, T-wave findings on antero-lateral,
postero-inferior, or antero-septal recording, Ventricular conduction
defect, White blood cell count

\hypertarget{final-expert-dag-with-added-nodes}{%
\subsection{Final Expert DAG With Added
Nodes}\label{final-expert-dag-with-added-nodes}}

Confounders measured at baseline: Age at entry, Alcohol use,
Antiarrhythmic agents, Antihypertensives other than diuretics, Atrial
fibrillation, Cognitive status, Depression, Diabetes, Digitalis,
Diuretics, Dyslipidemia, Education level, Employment status, History of
acute coronary insufficiency, History of angina pectoris, History of
congestive heart failure, History of intermittent cerebral ischemic
attack, History of intermittent claudication, Hypertension, Number of
myocardial infarctions at baseline, Occupation, Oral hypoglycemic
agents, Pre-randomization adherence, Race, Relative body weight, Risk
group, Time since most recent myocardial infarction

Counfounders measured at follow-up: Alcohol use, Antiarrhythmic agents,
Antihypertensives other than diuretics, Atrial fibrillation,
Cardiomegaly on chest X-ray, Cigarette smoking, Cognitive status,
Current habitual level of physical activity, Depression, Diabetes,
Digitalis, Diuretics, Dyslipidemia, Employment status, History of acute
coronary insufficiency, History of angina pectoris, History of
congestive heart failure, History of intermittent cerebral ischemic
attack, History of intermittent claudication, Hypertension, New York
Heart Association functional class, Occupation, Oral hypoglycemic
agents, Plasma fasting glucose

\hypertarget{non-confounders}{%
\subsection{Non-Confounders}\label{non-confounders}}

Confounders measured at baseline: Acute cholecystitis, Alpha-lipoprotein
cholesterol, Black stools, Blood sample turbidity, Blood type, Breast
enlargement, Bronchopulmonary abnormalitites, Burning when urinating,
Cause of death, Chemical composition of gallbladder stones, Chronic
cholecystitis, Cigarettes per day, Date of study entry, Decreased
appetite, Decreased libido, Dermatologic ailment, Difficulty swallowing,
Enlarged prostate, Enlarged thyroid, Enrolled in follow-up study, Eye
opacities, Far-sightedness, Finger tremor, Flushing, Forgetfulness,
Frequent urination, Gallbladder cancer, Gouty arthritis, Gynecomastia,
Hair loss, Heat intolerance, History of acute cholecystitis, History of
common duct stone, Increased appetite, Insulin use, Itching,
Musculoskeletal abnormalities, Nervous system abnormalities, Number of
kids, Palpable liver, Palpable spleen, Rales, Reduced flow of urine,
Return of unused study medication, Serum protein-bound iodine,
Short-sightedness, Shortness of breath at night, Sleeplessness, Stomach
pain, Supraventricular premature beats, Sweating, Swelling ankles,
Thrombophlebitis, Thyroid nodules, Ulcer, Unusually warm or moist skin,
Urinary stones, Urine conjugate of clofibrate, Urticaria, Vascular
spiders

\hypertarget{appendix-b-full-prompts}{%
\section*{Appendix B: Full Prompts}\label{appendix-b-full-prompts}}
\addcontentsline{toc}{section}{Appendix B: Full Prompts}

In each of the below prompts, \texttt{\{confounder\}} is replaced in the
prompt by the name of a potential confounder, which will often include a
measurement period indication, like ``baseline alcohol use'' or
``follow-up alcohol use.''

Prompt for the direct method, in which the LLM is directly asked whether
something is a confounder, with reasoning encouraged. There is an
alternate version of this prompt that is exactly the same except the
order of the options is changed to ``A. Not sure. B. Not a confounding
variable. C. A confounding variable.'':

\begin{quote}
You are a bot that specializes in causal inference in medical contexts.
When estimating the effect of X on Y in non-experimental data, you know
which variables are confounding factors that need to be accounted for to
isolate a causal effect.

I am going to ask you whether a specific variable is a confounder in my
analysis.

Example tasks and correct answers:

I have a data set consisting of a large representative sample of women.
I want to use causal analysis to figure out whether (hormone replacement
therapy) has a causal effect on (coronary artery disease).

In this analysis, the variable ``socioeconomic status'' is\ldots{}

A. Not a confounding variable B. A confounding variable C. Not sure.

CORRECT RESPONSE: Women who take hormone replacement therapy tend to be
wealthier and have access to better medical care, which will reduce the
incidence of coronary artery disease, so socioeconomic status is a
confounding variable. B.

I have a data set consisting of people over the age of 55. I want to use
causal analysis to figure out whether (ACE inhibitors) have a causal
effect on (myocardial infarction).

In this analysis, the variable ``takes daily low-dose aspirin''
is\ldots{}

A. Not a confounding variable B. A confounding variable C. Not sure.

CORRECT RESPONSE: People with more acute heart problems are often
recommended both ACE inhibitors and low-dose aspirin, so the two drugs
are correlated. Low-dose reduces mycardial infarction, so low-dose
aspirin is a confounding variable. B.

I have a data set of people who suffer from allergies. I want to use
causal analysis to figure out whether (taking acetaminophen) causally
reduces (cephalalgia / headache).

In this analysis, the variable ``takes regular vitamin D supplement''
is\ldots{}

A. Not a confounding variable B. A confounding variable C. Not sure.

CORRECT RESPONSE: Usage of acetaminophen and vitamin D may be
correlated. However, vitamin D supplementation should not affect
headache, so vitamin D is not a confounder in this analysis. A.

MY QUESTION FOR YOU:

I have a data set consisting only of people who have been assigned to
take a certain medication X. However, some of the patients choose to
take X as assigned, and others do not. I want to use causal analysis to
figure out whether (a patient's decision to take X) has a causal effect
on (the patient's mortality).

In this analysis, the variable (\{confounder\}) is\ldots{}

A. Not a confounding variable B. A confounding variable C. Not sure.

Explain your reasoning step-by-step, and then choose either A, B, or C.
The final word of your response should only be ``A.'', ``B.'', or
``C.''.
\end{quote}

Prompt for the indirect method, in which the LLM is separately asked
about the relationship between the confounder and adherence, and between
the confounder and mortality. The below prompt specifies (mortality),
but the alternate version is the exact same but replacing (mortality)
with (a patient's decision to take X).

\begin{quote}
You are a bot that specializes in causal inference in medical contexts.
You are familiar with the medical literature on whether a given variable
X is understood to have a causal effect on another variable Y.

I am going to ask you whether a specific variable is a cause of another
variable or not.

Example tasks and correct answers:

I have a data set consisting of a large representative sample of women.
Does (socioeconomic status) have a causal effect on (hormone replacement
therapy)?

A. Yes. (socioeconomic status) has a causal effect on (hormone
replacement therapy). B. No.~(socioeconomic status) and (hormone
replacement therapy) are statistically related, but there is not a
causal relationship between them in either direction. C.
No.~(socioeconomic status) and (hormone replacement therapy) are not
statistically or causally related to each other. D. Unsure.

CORRECT RESPONSE: A higher socioeconomic status improves access to
medical care, including hormone replacement therapy, which is expensive.
Socioeconomic status is a cause of hormone replacement therapy. A.

I have a data set consisting of people over the age of 55. Does (takes
daily low-dose aspirin) cause (takes ACE inhibitors)?

A. Yes. (takes daily low-dose aspirin) has a causal effect on (ACE
inhibitors). B. No.~(takes daily low-dose aspirin) and (ACE inhibitors)
are statistically related, but there is not a causal relationship
between them in either direction. C. No.~(takes daily low-dose aspirin)
and (ACE inhibitors) are not statistically or causally related to each
other. D. Unsure.

CORRECT RESPONSE: People with more acute heart conditions are likely to
both be recommended aspirin and ACE inhibitors. This will lead to a
correlation between the two. But it is the heart condition causing the
patient to take both. Taking aspirin does not cause someone to take ACE
inhibitors. B.

I have a data set of people who suffer from allergies. Does (takes
regular vitamin D supplement) causally reduce (cephalalgia / headache)?

In this analysis, the variable ``takes regular vitamin D supplement''
is\ldots{}

A. Yes. (takes regular vitamin D supplement) has a causal effect on
(cephalalgia / headache). B. No.~(takes regular vitamin D supplement)
and (cephalalgia / headache) are statistically related, but there is not
a causal relationship between them in either direction. C. No.~(takes
regular vitamin D supplement) and (cephalalgia / headache) are not
statistically or causally related to each other. D. Unsure.

CORRECT RESPONSE: There is no reason to believe that taking a vitamin D
supplement should be related to headache, causally or otherwise. C.

MY QUESTION FOR YOU:

I have a data set consisting only of people who have been assigned to
take a certain medication X. However, some of the patients choose to
take X as assigned, and others do not. In this data, does
(\{confounder\}) have a causal effect on (mortality)?

A. Yes. (\{confounder\}) has a causal effect on (mortality). B.
No.~(\{confounder\}) and (mortality) are statistically related, but
there is not a causal relationship between them in either direction. C.
No.~(\{confounder\}) and (mortality) are not statistically or causally
related to each other. D. Unsure.

Explain your reasoning step-by-step, and then choose either A, B, C, or
D. The final word of your response should only be ``A.'', ``B.'',
``C.'', or ``D.''.
\end{quote}

Prompt for the direct method with no reasoning encouraged:

\begin{quote}
You are a bot that specializes in causal inference in medical contexts.
When estimating the effect of X on Y in non-experimental data, you know
which variables are confounding factors that need to be accounted for to
isolate a causal effect.

I am going to ask you whether a specific variable is a confounder in my
analysis.

I have a data set consisting only of people who have been assigned to
take a certain medication X. However, some of the patients choose to
take X as assigned, and others do not. I want to use causal analysis to
figure out whether (a patient's decision to take X) has a causal effect
on (the patient's mortality).

In this analysis, the variable (\{confounder\}) is\ldots{}

A. Not a confounding variable B. A confounding variable C. Not sure.

Tell me your answer: A, B, or C. Give me ONLY your answer and no other
text in your response. The entire response should only be ``A.'',
``B.'', or ``C.''.
\end{quote}

Prompt for the indirect method with no reasoning encouraged:

\begin{quote}
You are a bot that specializes in causal inference in medical contexts.
You are familiar with the medical literature on whether a given variable
X is understood to have a causal effect on another variable Y.

I am going to ask you whether a specific variable is a cause of another
variable or not.

I have a data set consisting only of people who have been assigned to
take a certain medication X. However, some of the patients choose to
take X as assigned, and others do not. In this data, does
(\{confounder\}) have a causal effect on (mortality)?

A. Yes. (\{confounder\}) has a causal effect on (mortality). B.
No.~(\{confounder\}) and (mortality) are statistically related, but
there is not a causal relationship between them in either direction. C.
No.~(\{confounder\}) and (mortality) are not statistically or causally
related to each other. D. Unsure.

Tell me your answer: A, B, C, or D. Give me ONLY your answer and no
other text in your response. The entire response should only be ``A.'',
``B.'', ``C.'', or ``D.''.
\end{quote}

\hypertarget{appendix-c-reasons-for-non-confounder-status}{%
\section*{Appendix C: Reasons for Non-Confounder
Status}\label{appendix-c-reasons-for-non-confounder-status}}
\addcontentsline{toc}{section}{Appendix C: Reasons for Non-Confounder
Status}

Administrative study data: Blood sample turbidity, Cause of death, Date
of study entry, Enrolled in follow-up study, Return of unused study
medication

Anticipated side-effect or known metabolite of active study
medication(s): Alpha-lipoprotein cholesterol, Black stools, Breast
enlargement, Burning when urinating, Decreased appetite, Decreased
libido, Difficulty swallowing, Finger tremor, Flushing, Forgetfulness,
Frequent urination, Gynecomastia, Hair loss, Heat intolerance, Increased
appetite, Itching, Reduced flow of urine, Shortness of breath at night,
Sleeplessness, Stomach pain, Sweating, Swelling ankles, Ulcer, Urine
conjugate of clofibrate, Urticaria

General medical examination results: Blood type, Bronchopulmonary
abnormalitites, Enlarged prostate, Enlarged thyroid, Eye opacities,
Far-sightedness, Finger tremor, Gouty arthritis, Gynecomastia, Hair
loss, History of acute cholecystitis, History of common duct stone,
Musculoskeletal abnormalities, Nervous system abnormalities, Number of
kids, Palpable liver, Palpable spleen, Rales, Short-sightedness,
Thrombophlebitis, Thyroid nodules, Ulcer, Unusually warm or moist skin,
Urinary stones

Sub-study data collected only on a sample of participants: Acute
cholecystitis, Alpha-lipoprotein cholecystitis, Blood sample turbidity,
Chemical composition of gallbladder stones, Chronic cholecystitis,
Cigarettes per day, Gallbladder cancer, Insulin use, Serum protein-bound
iodine, Supraventricular premature beats, Ulcer, Urine conjugate of
clofibrate

\hypertarget{references}{%
\section*{References}\label{references}}
\addcontentsline{toc}{section}{References}

\hypertarget{refs}{}
\begin{CSLReferences}{1}{0}
\leavevmode\vadjust pre{\hypertarget{ref-anthropic_python}{}}%
Anthropic. 2024. {``{Anthropic Python API}.''}
\url{https://github.com/anthropics/anthropic-sdk-python}.

\leavevmode\vadjust pre{\hypertarget{ref-ashwani2024}{}}%
Ashwani, Swagata, Kshiteesh Hegde, Nishith Reddy Mannuru, Mayank Jindal,
Dushyant Singh Sengar, Krishna Chaitanya Rao Kathala, Dishant Banga,
Vinija Jain, and Aman Chadha. 2024. {``Cause and Effect: Can Large
Language Models Truly Understand Causality?''}
\url{https://doi.org/10.48550/ARXIV.2402.18139}.

\leavevmode\vadjust pre{\hypertarget{ref-cai2024}{}}%
Cai, Hengrui, Shengjie Liu, and Rui Song. 2024. {``Is Knowledge All
Large Language Models Needed for Causal Reasoning?''}
\url{https://doi.org/10.48550/ARXIV.2401.00139}.

\leavevmode\vadjust pre{\hypertarget{ref-cataloguebias}{}}%
Catalogue of Bias Collaboration, David Nunan, Jeffrey Aronson, and Clare
Bankhead. 2018. {``Confounding.''}
www.catalogueofbiases.org/biases/confounding.

\leavevmode\vadjust pre{\hypertarget{ref-influenc1980}{}}%
CDPRG. 1980. {``Influence of Adherence to Treatment and Response of
Cholesterol on Mortality in the Coronary Drug Project.''} \emph{New
England Journal of Medicine} 303 (18): 1038--41.
\url{https://doi.org/10.1056/nejm198010303031804}.

\leavevmode\vadjust pre{\hypertarget{ref-debertin2024}{}}%
Debertin, Julia, Javier A. Jurado Vélez, Laura Corlin, Bertha Hidalgo,
and Eleanor J. Murray. 2024. {``Synthesizing Subject-Matter Expertise
for Variable Selection in Causal Effect Estimation: A Case Study.''}
\emph{Epidemiology} 35 (5): 642--53.
\url{https://doi.org/10.1097/ede.0000000000001758}.

\leavevmode\vadjust pre{\hypertarget{ref-gururaghavendran2024}{}}%
Gururaghavendran, Rajesh, and Eleanor J. Murray. 2024. {``Can Algorithms
Replace Expert Knowledge for Causal Inference? A Case Study on Novice
Use of Causal Discovery.''} \emph{American Journal of Epidemiology},
August. \url{https://doi.org/10.1093/aje/kwae338}.

\leavevmode\vadjust pre{\hypertarget{ref-han2024}{}}%
Han, Sukjin. 2024. {``Mining Causality: AI-Assisted Search for
Instrumental Variables.''}
\url{https://doi.org/10.48550/ARXIV.2409.14202}.

\leavevmode\vadjust pre{\hypertarget{ref-humphrey2002postmenopausal}{}}%
Humphrey, Linda L, Benjamin KS Chan, and Harold C Sox. 2002.
{``Postmenopausal Hormone Replacement Therapy and the Primary Prevention
of Cardiovascular Disease.''} \emph{Annals of Internal Medicine} 137
(4): 273--84.

\leavevmode\vadjust pre{\hypertarget{ref-NEURIPS2023_631bb943}{}}%
Jin, Zhijing, Yuen Chen, Felix Leeb, Luigi Gresele, Ojasv Kamal, Zhiheng
LYU, Kevin Blin, et al. 2023. {``CLadder: Assessing Causal Reasoning in
Language Models.''} In \emph{Advances in Neural Information Processing
Systems}, edited by A. Oh, T. Naumann, A. Globerson, K. Saenko, M.
Hardt, and S. Levine, 36:31038--65. Curran Associates, Inc.
\url{https://proceedings.neurips.cc/paper_files/paper/2023/file/631bb9434d718ea309af82566347d607-Paper-Conference.pdf}.

\leavevmode\vadjust pre{\hypertarget{ref-kiciman2023}{}}%
Kıcıman, Emre, Robert Ness, Amit Sharma, and Chenhao Tan. 2023.
{``Causal Reasoning and Large Language Models: Opening a New Frontier
for Causality.''} \url{https://doi.org/10.48550/ARXIV.2305.00050}.

\leavevmode\vadjust pre{\hypertarget{ref-liu2024}{}}%
Liu, Xiaoyu, Paiheng Xu, Junda Wu, Jiaxin Yuan, Yifan Yang, Yuhang Zhou,
Fuxiao Liu, et al. 2024. {``Large Language Models and Causal Inference
in Collaboration: A Comprehensive Survey.''}
\url{https://doi.org/10.48550/ARXIV.2403.09606}.

\leavevmode\vadjust pre{\hypertarget{ref-long2023}{}}%
Long, Stephanie, Tibor Schuster, and Alexandre Piché. 2023. {``Can Large
Language Models Build Causal Graphs?''}
\url{https://doi.org/10.48550/ARXIV.2303.05279}.

\leavevmode\vadjust pre{\hypertarget{ref-murray2016}{}}%
Murray, Eleanor J., and Miguel A. Hernán. 2016. {``Adherence Adjustment
in the Coronary Drug Project: A Call for Better Per-Protocol Effect
Estimates in Randomized Trials.''} \emph{Clinical Trials} 13 (4):
372--78. \url{https://doi.org/10.1177/1740774516634335}.

\leavevmode\vadjust pre{\hypertarget{ref-murray2018}{}}%
---------. 2018. {``Improved Adherence Adjustment in the Coronary Drug
Project.''} \emph{Trials} 19 (1).
\url{https://doi.org/10.1186/s13063-018-2519-5}.

\leavevmode\vadjust pre{\hypertarget{ref-biolincc}{}}%
NHLBI BioLINCC. 2024. {``{Coronary Drug Project (CDP)}.''}
https://biolincc.nhlbi.nih.gov/studies/cdp/.

\leavevmode\vadjust pre{\hypertarget{ref-nori2023}{}}%
Nori, Harsha, Yin Tat Lee, Sheng Zhang, Dean Carignan, Richard Edgar,
Nicolo Fusi, Nicholas King, et al. 2023. {``Can Generalist Foundation
Models Outcompete Special-Purpose Tuning? Case Study in Medicine.''}
\url{https://doi.org/10.48550/ARXIV.2311.16452}.

\leavevmode\vadjust pre{\hypertarget{ref-openai_python}{}}%
OpenAI. 2024. {``{OpenAI Python API}.''}
\url{https://github.com/openai/openai-python}.

\leavevmode\vadjust pre{\hypertarget{ref-sheth2024}{}}%
Sheth, Ivaxi, Sahar Abdelnabi, and Mario Fritz. 2024. {``Hypothesizing
Missing Causal Variables with LLMs.''}
\url{https://doi.org/10.48550/ARXIV.2409.02604}.

\leavevmode\vadjust pre{\hypertarget{ref-takayama2024}{}}%
Takayama, Masayuki, Tadahisa Okuda, Thong Pham, Tatsuyoshi Ikenoue,
Shingo Fukuma, Shohei Shimizu, and Akiyoshi Sannai. 2024. {``Integrating
Large Language Models in Causal Discovery: A Statistical Causal
Approach.''} \url{https://doi.org/10.48550/ARXIV.2402.01454}.

\leavevmode\vadjust pre{\hypertarget{ref-zecevic2023}{}}%
Zečević, Matej, Moritz Willig, Devendra Singh Dhami, and Kristian
Kersting. 2023. {``Causal Parrots: Large Language Models May Talk
Causality but Are Not Causal.''} \emph{arXiv}.
\url{https://doi.org/10.48550/ARXIV.2308.13067}.

\leavevmode\vadjust pre{\hypertarget{ref-zhang2024}{}}%
Zhang, Huan, Yu Song, Ziyu Hou, Santiago Miret, and Bang Liu. 2024.
{``HoneyComb: A Flexible LLM-Based Agent System for Materials
Science.''} \url{https://doi.org/10.48550/ARXIV.2409.00135}.

\leavevmode\vadjust pre{\hypertarget{ref-zhou2023}{}}%
Zhou, Hongjian, Fenglin Liu, Boyang Gu, Xinyu Zou, Jinfa Huang, Jinge
Wu, Yiru Li, et al. 2023. {``A Survey of Large Language Models in
Medicine: Progress, Application, and Challenge.''}
\url{https://doi.org/10.48550/ARXIV.2311.05112}.

\end{CSLReferences}

\end{document}